\documentclass[aps,showpacs,showkeys,preprintnumbers,amsmath, amssymb,nofootinbib]{revtex4}
\usepackage{float}
\usepackage{graphics,epsfig}
\usepackage{graphicx}
\usepackage{dcolumn}
\usepackage{bm}

\begin{document}
\baselineskip=0.8 cm

\title{Towards experimentally studying some puzzles of Hawking radiation}
\author{Zehua Tian and Jiliang {Jing}\footnote{Electronic address:
jljing@hunnu.edu.cn}}

\affiliation{Department of Physics, and Key Laboratory of Low Dimensional Quantum Structures and
Quantum Control of Ministry of Education, Hunan Normal University,
Changsha, Hunan 410081, P. R. China}

\vspace*{0.2cm}
\begin{abstract}
\baselineskip=0.6 cm
\begin{center}
{\bf Abstract}
\end{center}

We investigate the features of the non-corrected thermal (non-thermal) spectrum and the quantum corrected thermal (non-thermal) spectrum. We find that: (i) using the quantum corrected non-thermal spectra, the black hole radiation as tunneling is an entropy conservation process, and thus black hole evaporation process is unitary; (ii) there are no obvious differences between all spectra except for near the Planck mass scale by comparing their average emission energies, average numbers of emissions and average emission energy fluctuations;  (iii) the energy covariances of Hawking radiations for all the thermal spectra are exactly zero, while they are nontrivial for all the non-thermal spectra. Especially, there are distinctly different maximums of energy covariances for the temperature-corrected and energy-corrected non-thermal spectra. Consequently, these differences provide a possible way towards experimentally analyzing whether the radiation spectrum of black hole is thermal or non-thermal with or without high order quantum corrections.

\end{abstract}

\pacs{04.70.Dy, 03.67.-a, 97.60.Lf}

\keywords{Hawking radiation spectra, quantum corrections and information loss paradox.}

\maketitle
\newpage

\section{Introduction}

A black hole, in quantum sense, is not completely black, which can radiate energies continuously. Furthermore, like the black body radiation, it has a temperature $T=\frac{\kappa}{2\pi}$, where $\kappa$ is the surface gravity of the black hole \cite{Hawking1,Hawking2}. However, for the thermal radiation spectrum \cite{Unruh,Page}, the so called ``information loss paradox" arises, i.e., if the radiation is thermal, there are no correlations between the emitted field quanta, and then one will lose information about the nature of the matter that originally formed the black hole. More technically, the complete evaporation of a black hole, whereby a pure quantum state evolves into a thermal state, would violate the quantum mechanical unitarity. In this regard, it is worthy to note that many groups \cite{Krauss,Bekenstein,Horowitz,Hawking3,Belokolos} have attempted addressing this puzzle, but none has been successful. Recently, Parikh and Wilczek \cite{Parikh} pointed out that Hawking radiation is completely non-thermal if one enforces the energy conservation law. Based on this idea, Zhang and Cai et al. \cite{zhang1} proved that for the non-thermal spectrum there are correlations between the radiations, and thus a queue of corrected radiations can transmit encoded information. According to careful calculations of entropy taken out by the emitted particles, they found that the black hole as tunneling is an entropy conservation process. Thus, they concluded that the black hole evaporation process is unitary. After their work, a lot of papers \cite{zhang2,zhang3,zhang4,Nozari1,Yi,Israel} extended this method to other backgrounds, such as quantum corrected black hole \cite{Yi} and non-commutated black hole \cite{zhang4}. Of great interest, all the reexaminations arrived at the same conclusion that information leaks out through the radiation, and the total entropy is conserved. So, their method may provide a possibility to solve the ``information loss paradox". Therefore, it is worthy to check whether the radiation spectrum is non-thermal or not.

On the other hand, based on the Hamilton-Jacobi method beyond semiclassical approximation, R. Banerjee and B. R. Majhi \cite{Banerjee} computed all quantum corrections in the single particle action revealing that these are proportional to the usual semiclassical contribution. They gave a quantum modified Hawking temperature and entropy. After their work, many papers appeared to discuss the quantum corrections to temperature and entropy for different backgrounds \cite{Lin1,Zhu,Bibhas,Modak,Jiang1,Jiang2,Mirza}. Especially, for this quantum corrected case, D. Singleton et al. \cite{Singleton} used the same method introduced in Ref. \cite{zhang1} to study the ``information loss paradox", and they also found that the total information will be carried away by the correlations of the outgoing radiations when the black hole evaporates completely. However, some authors \cite{Wang,Alexandre} argued that there are no quantum corrections to the Hawking temperature and entropy if one chooses the standard definition of the particle energy. And all the corrections are found to be the contributions to the particle energy. This choice is reasonable because it keeps the validity of the first law of the thermodynamics. Thus, a dispute, whether the Hawking temperature and entropy are modified or not, is caused.

It is worthy to note that Zhang et al. have compared the non-corrected thermal and non-thermal spectra, and they found the thermal spectrum can be distinguished with the non-thermal spectrum from their distinctly different energy covariances \cite{zhang5}. However, as introduced above, the puzzles of Hawking radiation are not just whether the radiation spectrum is thermal or not, they also contains other cases, such as whether the spectrum is quantum corrected or not, even the quantum corrected spectra, if there are quantum corrections to spectrum, is temperature-corrected or not. So, it is needed to consider all possible radiation spectra, and give a complete analysis. In this paper, we focus our attentions on these puzzles of Hawking radiation introduced above, and try to find out whether the radiation spectrum is thermal or non-thermal with or without high order quantum corrections.

Our paper is constructed as follows: In section \ref{section 1}, we simply discuss different radiation spectra coming from taking account of different conditions. In section \ref{section 2}, we try to analyze the ``information loss paradox" based on the temperature-corrected and the energy-corrected non-thermal spectra. In section \ref{section 3}, we compare the average emission energies, average numbers of emissions, and average emission energy fluctuations for the non-corrected thermal spectrum and quantum corrected thermal spectra.  In section \ref{section 4}, we compare the average emission energies {\it{et al.}} for the temperature-corrected thermal (non-thermal) spectrum with that for the energy-corrected thermal (non-thermal) spectrum. In section \ref{section 5} we give an analysis of energy covariances for the thermal spectra and non-thermal spectra. We finally summarize our conclusions in section \ref{section 6}.

\section{Introduction about different radiation spectra}\label{section 1}

There are six kinds of radiation spectra for black hole, i.e., non-corrected thermal and non-thermal spectra, temperature-corrected thermal and non-thermal spectra, and energy-corrected thermal and non-thermal spectra. We now listed them in the following.

\subsection{Non-corrected thermal and non-thermal spectra}

For the Schwarzschild black hole, the first obtained spectrum \cite{Hawking1,Hawking2} is
\begin{eqnarray}\label{thermal spectrum}
\Gamma_\mathrm{T}\sim\exp\left[-\frac{8\pi\varepsilon\mathcal{M}}{\hbar}\right],
\end{eqnarray}
which denotes the possibility of radiating a particle with energy $\varepsilon$ from the black hole with the mass $\mathcal{M}$, and it is obviously a thermal spectrum.

On the other hand, considering the energy conservation law during the evaporation process, Parikh and Wilczek \cite{Parikh} gave a non-thermal spectrum expressed as
\begin{eqnarray}\label{non-thermal spectrum}
\Gamma_{\mathrm{NT}}\sim\exp\left[-\frac{8\pi\varepsilon}{\hbar}(\mathcal{M}-\frac{\varepsilon}{2})\right].
\end{eqnarray}
Obviously, if we omit $\varepsilon^2$ correction, Eq. (\ref{non-thermal spectrum}) will go back to Eq. (\ref{thermal spectrum}).

\subsection{Temperature-corrected thermal and non-thermal spectra}

Banerjee and Majhi \cite{Banerjee}, from the Hamilton-Jacobi method beyond semiclassical approximation, got an quantum corrected spectrum
\begin{eqnarray}\label{thermal high order quantum corrected spectrum}
\nonumber
\Gamma_\mathrm{TCT}&\sim&\exp\left[-\frac{8\pi\varepsilon\mathcal{M}}{\hbar}
\big(1+\sum_i\beta_i\frac{\hbar^i}{\mathcal{M}^{2i}}\big)\right]
\\
&=&\exp\left[-\frac{8\pi\varepsilon\mathcal{M}}{\hbar}
\big(1-\frac{\alpha\hbar}{\mathcal{M}^{2}}\big)^{-1}\right],
\end{eqnarray}
where we have taken $\beta_i=\alpha^i$ as Ref. \cite{Banerjee}. Let's note that this choice is consistent with the result by considering the one loop back reaction effects in the spacetime \cite{York,Lousto}. Moreover, the coefficient $\alpha$ is related to the trace anomaly. Using conformal field theory techniques, Fursaev et al. \cite{Fursaev}  showed that for the Schwarzschild black hole $\alpha$ is
\begin{eqnarray}\label{Alpha}
\alpha=-\frac{1}{360\pi}\big(-N_0-\frac{7}{4}N_{\frac{1}{2}}+13N_1+\frac{233}{4}N_{\frac{3}{2}}-212N_2\big),
\end{eqnarray}
where $N_s$ denotes the number of fields with spin `$s$'.

Analogously, by using the connection between the tunneling rate (\ref{thermal high order quantum corrected spectrum}) and the change in entropy given in Ref. \cite{Parikh}, one can also get the non-thermal spectrum with high order quantum corrections
\begin{eqnarray}\label{non-thermal high order quantum corrected spectrum}
\Gamma_\mathrm{TCNT}\sim\left(\frac{-\alpha+(\mathcal{M}-\varepsilon)^2/\hbar}{-\alpha+\mathcal{M}^2/\hbar}\right)^{4\pi\alpha}
\exp\left[-\frac{8\pi\varepsilon}{\hbar}\big(\mathcal{M}-\frac{\varepsilon}{2}\big)\right].
\end{eqnarray}
Eqs. (\ref{thermal high order quantum corrected spectrum}) and (\ref{non-thermal high order quantum corrected spectrum}) will respectively go back to Eqs. (\ref{thermal spectrum}) and (\ref{non-thermal spectrum}) when $\alpha=0$, i.e., when there are no quantum corrections.

\subsection{Energy-corrected thermal and non-thermal spectra}

In the above discussions, the authors in Ref. \cite{Banerjee} have assumed that the energy of emitted particles is defined by
\begin{eqnarray}\varepsilon=-\partial_tI_0,
\end{eqnarray}
in which $I_0$ is the action of emitted particle without quantum corrections. Therefore, using $\Gamma\sim e^{-\varepsilon/T'}$, they \cite{Banerjee} get a quantum corrected Hawking temperature by Hamilton-Jacobi method beyond semiclassical approximation, which is given by
\begin{eqnarray} T'=\frac{\hbar}{8\pi\mathcal{M}}
\big(1+\sum_i\beta_i\frac{\hbar^i}{\mathcal{M}^{2i}}\big)^{-1}.\end{eqnarray}
We call this the temperature-corrected case. It is worthy to note that their method has been discussed and extended widely \cite{Lin1,Zhu,Bibhas,Modak,Jiang1,Jiang2,Mirza}.

However, we think that, in these discussions, we should use the standard definition of particle energy in curved spacetime \cite{book1}
\begin{eqnarray}\label{standard definition of particle energy}
\varepsilon'=-P_\mu\xi^\mu=-\xi^\mu\partial_\mu I,
\end{eqnarray}
where $I$ is the action of the emitted particle and $\xi^\mu$ is a timelike Killing vector. By using the Hamilton-Jacobi method beyond semiclassical approximation, we find that the particle energy is $\varepsilon'=\varepsilon (1+\sum_i\beta_i\frac{\hbar^i}{\mathcal{M}^{2i}})$ and then the Hawking temperature is
\begin{eqnarray}\label{standard temperature}
T=\frac{\hbar}{8\pi \mathcal{M}}.
\end{eqnarray}
Obviously, it is the standard Hawking temperature. This implies that, provided we take the standard definition of the particle energy, the Hawking temperature is not modified by the quantum tunneling beyond semiclassical approximation. Besides, according to $S=\int\frac{1}{T}d\mathcal{M}$, the entropy can be obtained as
\begin{eqnarray}\label{entropy}
S_{BH}=\frac{4\pi\mathcal{M}^2}{\hbar},
\end{eqnarray}
which is not modified too. Because the particle energy $\varepsilon'$ in this case is different from $\varepsilon$, we call it the energy-corrected case.
For the energy-corrected case, the possibility of emitted particles for thermal spectrum can be expressed as
\begin{eqnarray}\label{energy correction thermal possibility}
\nonumber
\Gamma_\mathrm{ECT}&\sim&\exp\left[-\frac{8\pi\varepsilon\mathcal{M}}{\hbar}
\big(1+\sum_i\beta_i\frac{\hbar^i}{\mathcal{M}^{2i}}\big)\right]
\\
&=&\exp\left[-\frac{8\pi\mathcal{M}\varepsilon'}{\hbar}\right].
\end{eqnarray}

Considering the energy conservation law, we can also obtain the possibility of emitted particles for non-thermal spectrum expressed as
\begin{eqnarray}\label{energy correction non-thermal possibility}
\Gamma_\mathrm{ECNT}\sim\exp\left[-\frac{8\pi\varepsilon'}{\hbar}(\mathcal{M}-\frac{\varepsilon'}{2})\right].
\end{eqnarray}

Compared with the non-corrected thermal and non-thermal spectra (\ref{thermal spectrum}) and (\ref{non-thermal spectrum}), the energy-corrected thermal and non-thermal spectra (\ref{energy correction thermal possibility}) and (\ref{energy correction non-thermal possibility}) have the only difference, the radiation energy. For the former case it is $\varepsilon$, while for the latter case it is $\varepsilon'=\varepsilon\big(1+\sum_i\beta_i\frac{\hbar^i}{\mathcal{M}^{2i}}\big)$.

Above discussions show us that there are several different spectra for the Schwarzschild black hole. The non-thermal spectra result from the energy conservation law, while quantum corrections come from the Hamilton-Jacobi method beyond semiclassical approximation (considering all quantum corrections). Furthermore, for the quantum corrected cases, different energy definitions of radiation particle can result in different Hawking temperature and entropy. In this regard, it is interesting to note that both the temperature and entropy of black hole are not modified by the quantum tunneling beyond semiclassical approximation if we use the standard definition of the particle energy $\varepsilon'=-P_\mu\xi^\mu$.

For the Schwarzschild black hole, so many spectra seem to be possible. One may ask which spectrum is real. In the following, we will review how to solve the ``information loss paradox" based on the temperature-corrected non-thermal spectrum and the energy-corrected one, and then study the features of the six different spectra with the hope of distinguishing them.

\section{Solutions to information paradox of black hole}\label{section 2}

An obvious difference between the non-thermal spectra and thermal spectra is that sequential emissions are correlated for the non-thermal spectra, while that for the thermal spectra are not. What's more, these correlations can reveal where the black hole information goes. Through a careful counting of the entropy taken out by the emitted particles, the black hole radiation as tunneling is proved to be an entropy conservation process. Zhang, Cai, You and Zhan \cite{zhang1} have proved that, using the non-corrected non-thermal spectrum,  the information is leaking out through the radiations, and the black hole evaporation process is unitary. We will show that, using the temperature-corrected non-thermal spectrum and the energy-corrected one, the information is also leaking out through the radiations, and the black hole evaporation process is still unitary. These give possible resolutions to the ``information paradox" under specific conditions.

\subsection{Solution to information paradox through temperature-corrected non-thermal spectrum}

Considering all quantum corrections in the single particle action, together with choosing particle energy $\varepsilon=-\partial_tI_0$, for a black hole with mass $\mathcal{M}$ one can get a quantum corrected temperature
\begin{eqnarray}
T'=\frac{\hbar}{8\pi\mathcal{M}}\big(1-\frac{\alpha\hbar}{\mathcal{M}^{2}}\big),
\end{eqnarray}
and entropy
\begin{eqnarray}\label{SSSS}
S'_{\mathrm{BH}}=\int\frac{1}{T'}d\mathcal{M}=4\pi\frac{\mathcal{M}^2}{\hbar}+4\pi\alpha \ln\left[1-\frac{\mathcal{M}^2}{\alpha\hbar}\right].
\end{eqnarray}

As showed in Ref. \cite{zhang1}, the entropy of the first emission with an energy $\varepsilon_1$ from a black hole of mass $\mathcal{M}$ is
\begin{eqnarray}\label{entropy 1}
\nonumber
S_{\mathrm{TCNT}}(\varepsilon_1)&=&-\ln\Gamma_{\mathrm{TCNT}}(\varepsilon_1)
\\
&=&\frac{8\pi\varepsilon_1}{\hbar}\big(\mathcal{M}-\frac{\varepsilon_1}{2}\big)-
4\pi\alpha\ln\left[\frac{-\alpha+(\mathcal{M}-\varepsilon_1)^2/\hbar}{-\alpha+\mathcal{M}^2/\hbar}\right].
\end{eqnarray}
The conditional entropy of a second emission with an energy $\varepsilon_2$ after the $\varepsilon_1$ emission
is
\begin{eqnarray}\label{entropy 2}
\nonumber
S_{\mathrm{TCNT}}(\varepsilon_2\mid\varepsilon_1)&=&-\ln\Gamma_{\mathrm{TCNT}}(\varepsilon_2\mid\varepsilon_1)
\\
&=&\frac{8\pi\varepsilon_2}{\hbar}\big(\mathcal{M}-\varepsilon_1-\frac{\varepsilon_2}{2}\big)-
4\pi\alpha\ln\left[\frac{-\alpha+(\mathcal{M}-\varepsilon_1-\varepsilon_2)^2/\hbar}
{-\alpha+(\mathcal{M}-\varepsilon_1)^2/\hbar}\right].
\end{eqnarray}
Repeating the process, we will find that the entropy of a Hawking emission, at an energy $\varepsilon_i$, conditional on the earlier emissions labeled by $\varepsilon_1$, $\varepsilon_2$,..., and $\varepsilon_{i-1}$, is of
\begin{eqnarray}\label{entropy i}
\nonumber
S_{\mathrm{TCNT}}(\varepsilon_i\mid\varepsilon_1,\varepsilon_2,...,\varepsilon_{i-1})&=&
\frac{8\pi\varepsilon_i}{\hbar}\big(\mathcal{M}-\sum^{i-1}_{j=1}\varepsilon_j-\frac{\varepsilon_i}{2}\big)
\\
&&-4\pi\alpha\ln\left[\frac{-\alpha+(\mathcal{M}-\sum^{i-1}_{j=1}\varepsilon_j-\varepsilon_i)^2/\hbar}
{-\alpha+(\mathcal{M}-\sum^{i-1}_{j=1}\varepsilon_j)^2/\hbar}\right].
\end{eqnarray}
Then, we can calculate the total entropy taken away by the emissions ($\varepsilon_1$, $\varepsilon_2$,..., $\varepsilon_n$,) that exhaust the initial black hole ($\mathcal{M}=\sum^n_{i=1}\varepsilon_i$), which is
\begin{eqnarray}\label{the total entropy}
\nonumber
S_{\mathrm{TCNT}}(\varepsilon_1,\varepsilon_2,...,\varepsilon_n)&=&\sum^n_{i=1}
S_{\mathrm{TCNT}}(\varepsilon_i\mid\varepsilon_1,\varepsilon_2,...,\varepsilon_{i-1})
\\
&=&4\pi\frac{\mathcal{M}^2}{\hbar}+4\pi\alpha \ln\left[1-\frac{\mathcal{M}^2}{\alpha\hbar}\right]=S'_{\mathrm{BH}}.
\end{eqnarray}
Eq. (\ref{the total entropy}) suggests that the entropy of all emitted Hawking radiations is equal to the entropy of the black hole, which implies no information is lost in the process of Hawking radiation. However, we should point out that the entropy (\ref{the total entropy}) is not the entropy of the initial Schwarzschild black hole, $S_{\mathrm{BH}}=4\pi\mathcal{M}^2/\hbar.$

\subsection{Solution to information paradox through energy-corrected non-thermal spectrum}

From Eq. (\ref{entropy}), we know that the entropy, although by considering all quantum corrections in the single particle action, is not modified if we choose the standard definition of particle energy. For this case, the entropy of a Hawking emission, at an energy $\varepsilon'_i$, conditional on the earlier emissions labeled by $\varepsilon'_1$, $\varepsilon'_2$, ..., and $\varepsilon'_{i-1}$, is of
\begin{eqnarray}\label{entropy i1}
S_{\mathrm{ECNT}}(\varepsilon'_i\mid\varepsilon'_1,\varepsilon'_2,...,\varepsilon'_{i-1})=
\frac{8\pi\varepsilon'_i}{\hbar}\big(\mathcal{M}-\sum^{i-1}_{j=1}\varepsilon'_j-\frac{\varepsilon'_i}{2}\big).
\end{eqnarray}
Thus, after the black hole evaporates completely, the total entropy for a given sequence of emissions ($\varepsilon'_1$, $\varepsilon'_2$, ..., $\varepsilon'_n$) with $\mathcal{M}=\sum^n_{i=1}\varepsilon'_i$ is
\begin{eqnarray}\label{the total entropy1}
\nonumber
S_{\mathrm{ECNT}}(\varepsilon'_1,\varepsilon'_2,...,\varepsilon'_n)&=&\sum^n_{i=1}
S_{\mathrm{ECNT}}(\varepsilon'_i\mid\varepsilon'_1,\varepsilon'_2,...,\varepsilon'_{i-1})
\\
&=&\frac{4\pi\mathcal{M}^2}{\hbar}=S_{\mathrm{BH}}.
\end{eqnarray}
We can see from Eq. (\ref{the total entropy1}) that the entropy of all emitted Hawking radiation is equal to the entropy of the initial black hole, which implies no information is lost in the process of Hawking radiation.

Above discussions show us that both the temperature-corrected non-thermal spectrum and the energy-corrected one can provide a possible way to solve the information paradox.


\section{Features of thermal spectra with or without high order quantum corrections} \label{section 3}

Given a queue of emissions, we can calculate the average energies and the covariances of the emitted radiations for the different spectra: thermal and non-thermal with or without quantum corrections (temperature- or energy-corrected). These are studied below after introducing suitable units for discussing the Hawking radiations and their associated properties.

To prepare for numerical comparison of the observables associated with these spectra, we need to normalize them according to
\begin{eqnarray}\label{nomalization}
\int^{\mathcal{M}}_0\Gamma(\varepsilon)d\varepsilon=1.
\end{eqnarray}
Meanwhile, because the spectrum for Hawking radiation is a function of physical quantities from very large quantity ( $c$ ) to very small ones ($\hbar$ and $G$), a convenient modus is to introduce some dimensionless quantities, such as dimensionless mass $M=\mathcal{M}/M_P=\mathcal{M}/\sqrt{\hbar}$ and energy $E=\varepsilon/\kappa_bT_P=\varepsilon/\sqrt{\hbar}$. This is because $\hbar$, by assuming $G=c=\kappa_B=1$, is of the order of square of the Planck Mass $M_P$ \cite{Banerjee}. After doing this, we can rewrite the three thermal spectra introduced above in terms of dimensionless form as
\begin{eqnarray}\label{three thermal spectra1}
&& \mathrm{Non-corrected}:~~~~~~~~~~~~~~~ \Gamma_\mathrm{T}(E)=\frac{M\exp(-ME)}{1-\exp(-8\pi M^2)},
\\ \label{three thermal spectra2}
&& \mathrm{Temperature-corrected}: ~~~~ \Gamma_\mathrm{TCT}(E)=\frac{M^3\exp\left[-ME(1-\frac{\alpha}{M^2})^{-1}\right]}
{(M^2-\alpha)(1-\exp\big[\frac{8\pi M^4}{\alpha-M^2}\big])},
\\ \label{three thermal spectra3}
&& \mathrm{Energy-corrected}: ~~~~~~~~~~~~ \Gamma_\mathrm{ECT}(E)=\frac{M\exp(-ME(1-\frac{\alpha}{M^2})^{-1})}{1-\exp(-8\pi M^2)}.
\end{eqnarray}
where $E\in[0, 8\pi M]$.

\begin{figure}[ht]
\centering
\includegraphics[width=0.45\textwidth]{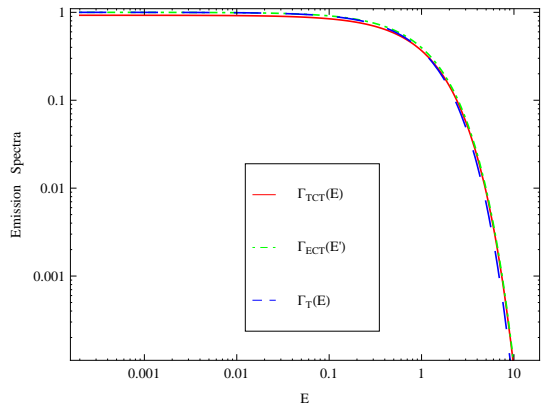}
\caption{(color online) The three thermal spectra are compared for a black hole at the Planck mass scale. The red solid, green dot-dashed and blue dashed lines respectively refer to the temperature-corrected, the energy-corrected and non-corrected thermal spectrum.}\label{f1}
\end{figure}

To make sure that the leading order correction to entropy is the same with that found earlier in Refs. \cite{Das,More} by a statistical method, we also take $\alpha=-\frac{1}{4\pi}$ hereafter. In Fig. \ref{f1} we compare the three thermal spectra (\ref{three thermal spectra1}), (\ref{three thermal spectra2}) and (\ref{three thermal spectra3}) for the Schwarzschild black hole at the Planck mass scale. It is shown that there are few difference between them, especial between the  temperature-corrected spectrum and the energy-corrected spectrum, they almost stay the same.

\subsection{Average energies for non-corrected, temperature-corrected and energy-corrected thermal spectra}

In units of $M_P$, we compute the average energy of Hawking radiations at any instant. For the thermal spectrum without corrections, we have
\begin{eqnarray}\label{average energy1}
\langle E(M)\rangle_{\mathrm{T}}=\left(M+\frac{8\pi M^3}{e^{8\pi M^2}-1-8\pi M^2}\right)^{-1}.
\end{eqnarray}
It is easy to find that $\langle E(M)\rangle_{\mathrm{T}}$ approaches $4\pi M$ for $M\ll1$ and $1/M$ for $M\gg1$.

For the temperature-corrected thermal spectrum, we find
\begin{eqnarray}\label{average energy2}
\langle E(M)\rangle_{\mathrm{TCT}}=\left(\frac{M^3(e^{\frac{8\pi M^4}{\alpha-M^2}}-1)}
{\alpha-M^2+e^{\frac{8\pi M^4}{\alpha-M^2}}(M^2-\alpha+8\pi M^4)}\right)^{-1}.
\end{eqnarray}
This quantity also approaches $4\pi M$ when $M\ll1$, and approaches $1/M$ when $M\gg1$.

For the energy-corrected thermal spectrum, the average energy is of
\begin{eqnarray}\label{average energy3}
\langle E'(M)\rangle_{\mathrm{ECT}}=\left(M+\frac{8\pi M^3}{e^{8\pi M^2}-1-8\pi M^2}\right)^{-1}.
\end{eqnarray}
It is interesting to note that it is the same with (\ref{average energy1}).

\begin{figure}[ht]
\centering
\includegraphics[width=0.46\textwidth]{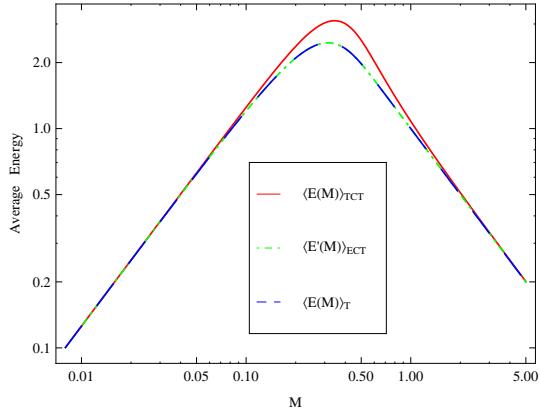}
\caption{(color online) The average emission energies for the temperature-corrected thermal spectrum (Red solid line), the energy-corrected thermal spectrum (Green dot-dashed line) and the non-corrected thermal spectrum (Blue dashed line).}\label{f2}
\end{figure}

We plot these three average energies of emitted particles in Fig. \ref{f2} as a function of the dimensionless mass $M$. It is shown that the average energy for the temperature-corrected case is noticeably different from that for the other two cases only near the Planck mass scale.

\subsection{Average number of radiations for non-corrected, temperature-corrected and energy-corrected thermal spectra}
For a black hole with mass $M$, the average number of radiations, according to the average energy (\ref{average energy1}), can be obtained for the non-corrected thermal spectrum approximately as
\begin{eqnarray}\label{average number1}
N_\mathrm{T}(M)=\frac{8\pi M}{\langle E(M)\rangle}_\mathrm{T}=\frac{8\pi M^2(e^{8\pi M^2}-1)}{e^{8\pi M^2}-1-8\pi M^2}.
\end{eqnarray}
We should pointed out that the result given in Ref. \cite{zhang5} may have a typos.

For the temperature-corrected thermal spectrum, we obtain the corresponding average number of radiations according to (\ref{average energy2}),
\begin{eqnarray}\label{average number2}
N_\mathrm{TCT}(M)=\frac{8\pi M}{\langle E(M)\rangle}_\mathrm{TCT}= \frac{8\pi M^4(e^{\frac{8\pi M^4}{\alpha-M^2}}-1)}
{\alpha-M^2+e^{\frac{8\pi M^4}{\alpha-M^2}}(M^2-\alpha+8\pi M^4)}.
\end{eqnarray}

Based on (\ref{average energy3}), for the energy-corrected thermal spectrum we have its corresponding average number of radiations
\begin{eqnarray}\label{average number3}
N_\mathrm{ECT}(M)=\frac{8\pi M}{\langle E'(M)\rangle_{\mathrm{ECT}}}
=\frac{8\pi M^2(e^{8\pi M^2}-1)}{e^{8\pi M^2}-1-8\pi M^2}.
\end{eqnarray}

It is interesting to note that the non-corrected case and the energy-corrected case give the same average number of radiations, this directly derives from their same average energies of emitted particles (\ref{average energy1}) and (\ref{average energy3}). Furthermore, Eqs. (\ref{average number1}), (\ref{average number2}) and (\ref{average number3}) give the same limits, $8\pi M^2$ for $M\gg1$ and $2$ for $M\ll1$.

\begin{figure}[ht]
\centering
\includegraphics[width=0.46\textwidth]{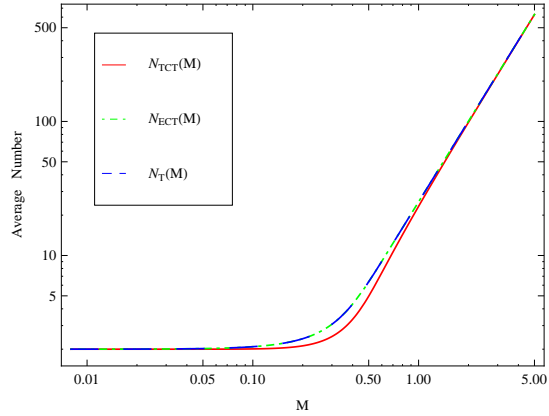}
\caption{(color online) The average number of emissions for the temperature-corrected thermal spectrum (Red solid line), the energy-corrected thermal spectrum (Green dot-dashed line) and the non-corrected thermal spectrum (Blue dashed line).}\label{f3}
\end{figure}

In Fig. \ref{f3}, we show the average number of emissions for the three thermal spectra. Above the Planck mass scale, the average number of emissions increases rapidly with the black hole mass. However, for the small Planck mass scale, it remains nearly a constant. Moreover, the average number of emissions for the temperature-corrected case almost has no difference with that of the other two cases.

\subsection{Standard deviations of emission energies for non-corrected, temperature-corrected and energy-corrected thermal spectra}
We now calculate the standard deviations of the emission energies, and we find for the non-corrected thermal spectrum
\begin{eqnarray}\label{standard deviation1}
\nonumber
\delta E^2_\mathrm{T}(M)&=&\langle E^2(M)\rangle_\mathrm{T}-\langle E(M)\rangle^2_\mathrm{T}
\\
&=&\frac{(\cosh8\pi M^2-1-32\pi^2M^2)\mathrm{csch}^24\pi M^2}{2M^2},
\end{eqnarray}
which has two limits, $16\pi^2M^2/3$ for $M\ll1$ and $1/M^2$ for $M\gg1$.

Analogously, for the temperature-corrected thermal spectrum, we find its corresponding standard deviations of the emission energies
\begin{eqnarray}\label{standard deviation2}
\delta E^2_\mathrm{TCT}(M)=(\frac{1}{M}-\frac{\alpha}{M^3})^2-64\pi^2M^2\frac{\exp(\frac{8\pi m^4}{\alpha-M^2})}{(\exp(\frac{8\pi m^4}{\alpha-M^2})-1)^2},
\end{eqnarray}
whose large and small $M$ limits are the same with that of Eq. (\ref{standard deviation1}).

For the energy-corrected thermal spectrum, the standard deviations of the emission energies is of
\begin{eqnarray}\label{standard deviation3}
\delta E'^2_\mathrm{ECT}(M)
=\frac{(\cosh8\pi M^2-1-32\pi^2M^2)\mathrm{csch}^24\pi M^2}{2M^2}.
\end{eqnarray}
It is worthy to note that the standard deviations (\ref{standard deviation3}) is the same with that for the non-corrected thermal spectrum described by Eq. (\ref{standard deviation1}).

\begin{figure}[ht]
\centering
\includegraphics[width=0.46\textwidth]{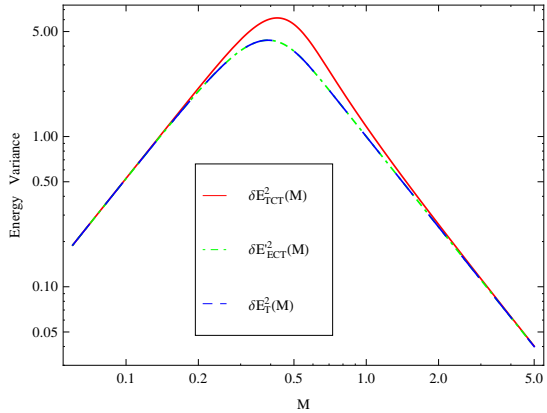}
\caption{(color online) The radiation energy variances for the temperature-corrected thermal spectrum (Red solid line), the energy-corrected thermal spectrum (Green dot-dashed line) and the non-corrected thermal spectrum (Blue dashed line).}\label{f4}
\end{figure}

Fig. \ref{f4} shows the standard deviations of the emission energies for the three cases (\ref{standard deviation1}), (\ref{standard deviation2}) and (\ref{standard deviation3}). Clearly, these features illustrates the dependence of the variances on the average energy.

From our analysis, we find no obvious differences between the temperature-corrected thermal spectrum and the non-corrected thermal spectrum (the energy-corrected thermal spectrum) except for tiny black holes with masses near the Planck mass scale. Especially, the three compared quantities for the energy-corrected thermal spectrum are completely the same with that for the non-corrected thermal spectrum.


\section{Comparing features of quantum corrected thermal and non-thermal spectra}\label{section 4}

Considering the energy conservation law, the radiation spectra is non-thermal \cite{Parikh}. This non-thermal feature shows that the emissions particles is correlated, and it provides a possible way to explain where the information goes during the evaporation process of black hole \cite{zhang1,zhang2}, while the thermal spectra can not. Therefore, it is significant to distinguish thermal spectrum and non-thermal spectrum. In the following, we will compare four quantum corrected spectra: the temperature-corrected thermal spectrum, the temperature-corrected non-thermal spectrum, the energy-corrected thermal spectrum and the energy-corrected non-thermal spectrum.

By using the dimensionless quantities $M$ and $E$, for the temperature-corrected non-thermal spectrum we find
\begin{eqnarray}\label{non-thermal spectrum1}
\Gamma_\mathrm{TCNT}=\Lambda(M)\big(\frac{-\alpha+(M-E/8\pi)}{-\alpha+M^2}\big)^{4\pi\alpha}
\exp\left[-E(M-\frac{E}{16\pi})\right],
\end{eqnarray}
where $\Lambda(M)=\left(\int^{8\pi M}_0\frac{\exp\left[-E(M-\frac{E}{16\pi})\right]}{\big(\frac{-\alpha+(M-E/8\pi)}
{-\alpha+M^2}\big)^{-4\pi\alpha}}dE\right)^{-1}$ is the normalization constant.

For the energy-corrected non-thermal spectrum, it is of
\begin{eqnarray}\label{non-thermal spectrum2}
\Gamma_\mathrm{ECNT}=\frac{1}{4\sqrt{\pi}\mathrm{F}[2\sqrt{\pi}M]}
\exp\left[-E(1-\frac{\alpha}{M^2})^{-1}\left(M-\frac{E}{16\pi}(1-\frac{\alpha}{M^2})^{-1}\right)\right],
\end{eqnarray}
where $\mathrm{F}[x]$ is Dawson function.

\begin{figure}[ht]
\centering
\includegraphics[width=0.46\textwidth]{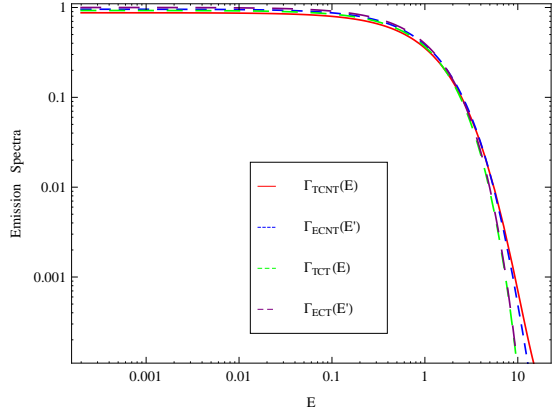}
\caption{(color online) The four spectra compared for a black hole at the Planck mass scale. The red solid, blue dashed, green dashed and purple dashed lines respectively refers to the temperature-corrected non-thermal, energy-corrected non-thermal, temperature-corrected thermal and energy-corrected thermal spectra.}\label{f5}
\end{figure}

Fig. \ref{f5} compares the non-thermal spectra (\ref{non-thermal spectrum1}) and (\ref{non-thermal spectrum2}) with the thermal spectra (\ref{three thermal spectra2}) and (\ref{three thermal spectra3}) for a black hole at the Planck mass scale. It is shown that the difference between the non-thermal spectra and the thermal spectra concentrated near $E\sim T_M/M^2_P$ with the equivalent Hawking radiation temperature $T_M$ for a black hole of mass $M$ measured in units of $M_P$.

\subsection{Average energies for temperature-corrected and energy-corrected non-thermal spectra}

In units of $M_P$, as for a fixed mass black hole, we calculate the average energy of Hawking emissions for the temperature-corrected non-thermal case, which is
\begin{eqnarray}\label{average energy4}
\langle E(M)\rangle_\mathrm{TCNT}=\int^{8\pi M}_0E\frac{\Lambda(M)\exp\left[-E(M-\frac{E}{16\pi})\right]}
{\big(\frac{-\alpha+(M-E/8\pi)}{-\alpha+M^2}\big)^{-4\pi\alpha}}dE.
\end{eqnarray}

For the energy-corrected non-thermal case, the corresponding average energy of Hawking emissions is
\begin{eqnarray}\label{average energy5}
\langle E'(M)\rangle_\mathrm{ECNT}=8\pi M-\frac{4(e^{4\pi M^2}-1)}{\mathrm{Er}[2\sqrt{\pi}M]},
\end{eqnarray}
where $\mathrm{Er}[x]$ is Error function. Interestingly, Eq. (\ref{average energy5}) has the same limits
with that of Eq. (\ref{average energy3}) for $M\ll1$ and $M\gg1$.

\begin{figure}[ht]
\centering
\includegraphics[width=0.46\textwidth]{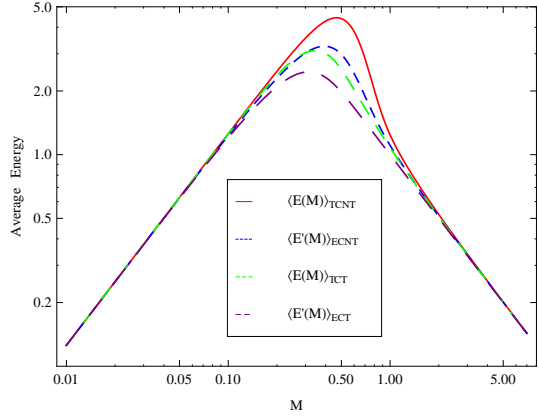}
\caption{(color online) The average emission energies for the temperature-corrected non-thermal spectrum (Red solid line), the energy-corrected non-thermal spectrum (Blue dashed line), the temperature-corrected thermal spectrum (Green dashed line) and the energy-corrected thermal spectrum (Purple dashed line).}\label{f6}
\end{figure}

We plot Eqs. (\ref{average energy2}), (\ref{average energy3}), (\ref{average energy4}) and (\ref{average energy5}) as a function of dimensionless mass $M$ in Fig. \ref{f6}. Which illustrates that the distinguishable feature between the non-thermal spectra and the thermal spectra is only near the Planck mass scale. Besides, it is also near the Planck mass scale where the temperature-corrected spectra and the energy-corrected spectra have the noticeable difference.

\subsection{Average number of radiations for temperature-corrected and energy-corrected non-thermal spectra}

According to the average energies Eqs. (\ref{average energy4}) and (\ref{average energy5}), it is easy to obtain the average number of radiations emitted from a black hole with mass $M$. For the temperature-corrected non-thermal spectrum, it is of
\begin{eqnarray}\label{average number4}
N_\mathrm{TCNT}(M)=\frac{8\pi M}{\langle E(M)\rangle_\mathrm{TCNT}}=\frac{8\pi M}{\int^{8\pi M}_0E\frac{\Lambda(M)\exp\left[-E(M-\frac{E}{16\pi})\right]}
{\big(\frac{-\alpha+(M-E/8\pi)}{-\alpha+M^2}\big)^{-4\pi\alpha}}dE}.
\end{eqnarray}
For the energy-corrected non-thermal spectrum, the average number of emitted particles is of
\begin{eqnarray}\label{average number5}
N_\mathrm{ECNT}(M)=\frac{2\pi M\mathrm{Er}[2\sqrt{\pi}M]}{2\pi M\mathrm{Er}[2\sqrt{\pi}M]-(e^{4\pi M^2}-1)}.
\end{eqnarray}
It approaches $2$ and $8\pi M^2$ for $M\ll1$ and $M\gg1$, respectively. And these limits are the same with that of Eqs. (\ref{average number2}) and (\ref{average number3}).

\begin{figure}[ht]
\centering
\includegraphics[width=0.46\textwidth]{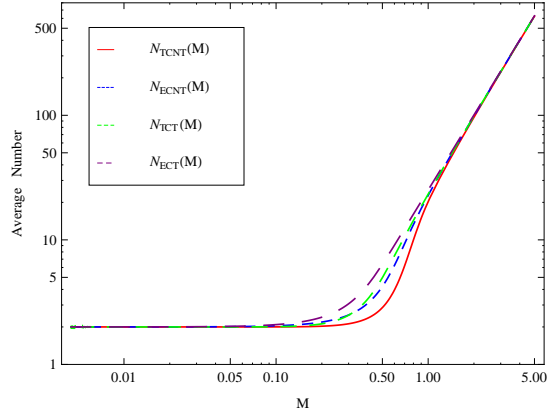}
\caption{(color online) The average number of emissions for the temperature-corrected non-thermal spectrum (Red solid line), the energy-corrected non-thermal spectrum (Blue dashed line), the temperature-corrected thermal spectrum (Green dashed line) and the energy-corrected thermal spectrum (Purple dashed line).}\label{f7}
\end{figure}

In Fig. \ref{f7}, we compare the average number of emissions (\ref{average number2}), (\ref{average number3}), (\ref{average number4}) and (\ref{average number5}). We find that for the small Planck mass scale the average number for all the four cases remains nearly a constant. Above the Planck mass scale, however, the average number of emissions increases rapidly with the increase of black hole mass. Furthermore, we can seen from Fig. \ref{f7} that average number of emissions for the four thermal spectra has no noticeable difference except for near the Planck mass scale.

\subsection{Standard deviations of emission energies for temperature-corrected and energy-corrected non-thermal spectra}
We now analyse the standard deviations of the emission energies. For the temperature-corrected
non-thermal spectrum, we find
\begin{eqnarray}\label{standard deviation4}
\nonumber
\delta E^2_\mathrm{TCNT}(M)&=&
\langle E^2(M)\rangle_\mathrm{TCNT}-\langle E(M)\rangle^2_\mathrm{TCNT}
\\\nonumber
&=&\int^{8\pi M}_0E^2\frac{\Lambda(M)\exp\left[-E(M-\frac{E}{16\pi})\right]}
{\big(\frac{-\alpha+(M-E/8\pi)}{-\alpha+M^2}\big)^{-4\pi\alpha}}dE
\\
&-&\left(\int^{8\pi M}_0E\frac{\Lambda(M)\exp\left[-E(M-\frac{E}{16\pi})\right]}
{\big(\frac{-\alpha+(M-E/8\pi)}{-\alpha+M^2}\big)^{-4\pi\alpha}}dE\right)^2.
\end{eqnarray}
And for the energy-corrected non-thermal spectrum, we find
\begin{eqnarray}\label{standard deviation5}
\delta E'^2_\mathrm{ECNT}(E)=\frac{8\pi\mathrm{Er}[2\sqrt{\pi}M]\big(4Me^{4\pi M^2}-\mathrm{Er}[2\sqrt{\pi}M]\big)-16\big(e^{4\pi M^2}-1\big)^2}{\mathrm{Er}^2[2\sqrt{\pi}M]}.
\end{eqnarray}
This quantity has two limits, $16\pi^2 M^2/3$ for $M\ll1$ and $1/M^2$ for $M\gg1$.

\begin{figure}[ht]
\centering
\includegraphics[width=0.46\textwidth]{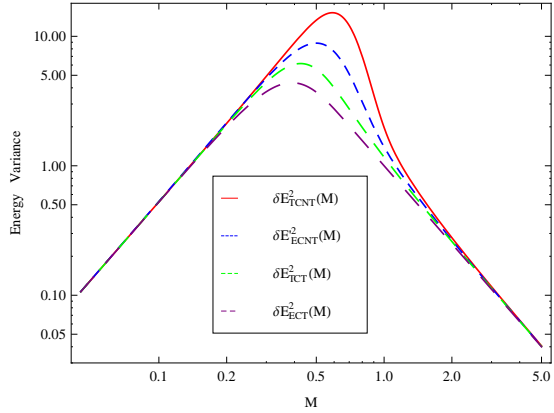}
\caption{(color online) The radiation energy variances for the temperature-corrected non-thermal spectrum (Red solid line), the energy-corrected non-thermal spectrum (Blue dashed line), the temperature-corrected thermal spectrum (Green dashed line) and the energy-corrected thermal spectrum (Purple dashed line).}\label{f8}
\end{figure}

Fig. \ref{f8} compares the Eqs. (\ref{standard deviation2}), (\ref{standard deviation3}), (\ref{standard deviation4}) and (\ref{standard deviation5}). Which tells us that the noticeable difference between the thermal and the non-thermal spectra exists only near the Planck mass scale. These features illustrates the dependence of the variance on the average energy.

According to our extensive analysis, we find that there is no drastic difference between the four spectra, the temperature-corrected thermal (non-thermal) spectrum, the energy-corrected thermal (non-thermal) spectrum, except for tiny black holes with masses near the Planck mass scale. Therefore, one would conclude that it is essentially impossible to experimentally distinguish the temperature-corrected thermal (non-thermal) spectrum from the energy-corrected thermal (non-thermal) one. And then the puzzles of Hawking radiation, whether information can be carried out from a black hole by correlations hidden in the emissions, whether the radiation spectrum has the high order quantum corrections, and the corrections are temperature-corrected or energy-corrected, still perplexes us. Nevertheless, we demonstrate below that information stored in the correlations of Hawking radiations from the non-thermal spectrum can indeed be observed through a counting of the emission energy covariances. Moreover, the emission energy covariances for the temperature-corrected non-thermal spectrum and the energy-corrected non-thermal spectrum have distinctly different maximums.


\section{Energy covariances}\label{section 5}
It is well known that for the thermal spectrum individual emissions are uncorrelated \cite{Hawking4}, and
one thus expects a vanishing covariance. Indeed, according to calculations, we obtain $\langle E_i(M)\rangle_{\mathrm{T}(\mathrm{TCT},\mathrm{ECT})}=\langle E_{j\neq i}(M)\rangle_{\mathrm{T}(\mathrm{TCT},\mathrm{ECT})}=\langle E(M)\rangle_{\mathrm{T}(\mathrm{TCT},\mathrm{ECT})}$ when individual emission energies are averaged over an ideal blackbody spectrum. So, we finally obtain
\begin{eqnarray}\label{covariance1}
\nonumber
\delta E^{2(\mathrm{cov})}_{\mathrm{T}(\mathrm{TCT}, \mathrm{ECT})}&=&
\langle E_i(M)E_{j\neq i}(M)\rangle_{\mathrm{T}(\mathrm{TCT}, \mathrm{ECT})}-
\langle E_i(M)\rangle_{\mathrm{T}(\mathrm{TCT}, \mathrm{ECT})}\langle E_{j\neq i}(M)\rangle_{\mathrm{T}(\mathrm{TCT}, \mathrm{ECT})}
\\
&=&0.
\end{eqnarray}

For the non-thermal spectra, we find that the average cross energy therm $\langle E_i(M)E_{j\neq i}(M)\rangle$, due to the existence of correlations between emissions, is nontrivial. And this quantity is strongly correlated with the probability for two emissions, one at an energy $E_i$ and another at an energy $E_j$. In this regard, let's note that the possibility satisfies  $\Gamma_\mathrm{NT}(E_1,E_2)=\Gamma_\mathrm{NT}(E_1+E_2)$ for an extensive list of black holes as shown in Ref. \cite{zhang1}. What's more, a recursive use of this relation allows us to show
\begin{eqnarray}\label{relation}
\Gamma_\mathrm{NT}(E_1,E_2)=\Gamma_\mathrm{NT}(E_1+E_2)=\Gamma_\mathrm{NT}(\widetilde{E}_1,E_1+E_2-\widetilde{E}_1),
\end{eqnarray}
as long as $E_1+E_2-\widetilde{E}_1>0$, or the probability for emissions $E_1, E_2, E_3,...$ is the same as the probability for the emission of a single radiation with an energy $\sum_jE_j$. Obviously, this probability distribution is symmetric with respect to any permutations of the individual emission indices. Thus it allows us to work within one sector and define the normalized probability subjected to the energy conservation constraint $\sum_jE_j\in[0,8\pi M]$.

According to the above analysis, for multiple emissions Eqs. (\ref{non-thermal spectrum1}) and (\ref{non-thermal spectrum2}) thus can be rewritten as
\begin{eqnarray}\label{spectrum for multiple emissions1}
\Gamma_\mathrm{TCNT}(\sum_jE_j)\sim\big(\frac{-\alpha+(M-\sum_jE_j/8\pi)}{-\alpha+M^2}\big)^{4\pi\alpha}
\exp\left[-\sum_jE_j(M-\frac{\sum_jE_j}{16\pi})\right],
\end{eqnarray}
and
\begin{eqnarray}\label{spectrum for multiple emissions2}
\Gamma_\mathrm{ECNT}(\sum_jE'_j)\sim\exp\left[-\sum_jE_j(1-\frac{\alpha}{M^2})^{-1}
\left(M-\frac{\sum_jE_j}{16\pi}(1-\frac{\alpha}{M^2})^{-1}\right)\right],
\end{eqnarray}
which is symmetric with respect to all permutations of indices. However, unlike Eqs. (\ref{non-thermal spectrum1}) and (\ref{non-thermal spectrum2}), we must normalize $\Gamma_\mathrm{TCNT(ECNT)}(E_1, E_2, E_3,...)$ according to
$\int^{8\pi M}_0dE_1\int^{8\pi M-E_1}_0dE_2...\Gamma_\mathrm{TCNT(ECNT)}(E_1, E_2, E_3,...)=1$.
Doing like this, for the energy-corrected non-thermal spectrum we finally give the normalization constant
\begin{eqnarray}
1/8\pi(-1+e^{-4\pi M^2}(1+2\pi M\mathrm{Er}[2\sqrt{\pi}M])),
\end{eqnarray}
for the case of two emissions with energies $E_1$ and $E_2$, and for the temperature-corrected non-thermal spectrum its normalization constant is of
\begin{eqnarray}
\Lambda_2(M)=\left(\int^{8\pi M}_0\int^{8\pi M-E_1}_0
\frac{\exp\left[-(E_1+E_2)(M-\frac{E_1+E_2}{16\pi})\right]}
{\big(\frac{-\alpha+(M-(E_1+E_2)/8\pi)}{-\alpha+M^2}\big)^{-4\pi\alpha}}dE_2dE_1\right)^{-1}.
\end{eqnarray}

\subsection{Energy covariance for temperature-corrected non-thermal spectrum}

The covariances of successive emissions for the temperature-corrected non-thermal spectrum can be obtained
\begin{eqnarray}
\nonumber
\delta E^{2(\mathrm{cov})}_\mathrm{TCNT}(M)&=&
\langle E_1(M)E_2(M)\rangle_\mathrm{TCNT}-
\langle E_1(M)\rangle_\mathrm{TCNT}\langle E_2(M)\rangle_\mathrm{TCNT}
\\ \nonumber
&=&\int^{8\pi M}_0\int^{8\pi M-E_1}_0E_1E_2
\frac{\Lambda_2(M)\exp\left[-(E_1+E_2)(M-\frac{E_1+E_2}{16\pi})\right]}
{\big(\frac{-\alpha+(M-(E_1+E_2)/8\pi)}{-\alpha+M^2}\big)^{-4\pi\alpha}}dE_2dE_1
\\ \nonumber
&&-\left(\int^{8\pi M}_0E_1\frac{\Lambda_(M)\exp\left[-E_1(M-\frac{E_1}{16\pi})\right]}
{\big(\frac{-\alpha+(M-E_1/8\pi)}{-\alpha+M^2}\big)^{-4\pi\alpha}}dE_1\right)
\\
&&\times\left(\int^{8\pi M}_0E_2\frac{\Lambda_(M)\exp\left[-E_2(M-\frac{E_2}{16\pi})\right]}
{\big(\frac{-\alpha+(M-E_2/8\pi)}{-\alpha+M^2}\big)^{-4\pi\alpha}}dE_2\right).
\end{eqnarray}

Here, it is needed to point out that we cannot get an analytical formula, so we give a numerical integral in the following figure.

\subsection{Energy covariance for energy-corrected non-thermal spectrum}

For the energy-corrected non-thermal spectrum, its corresponding energy covariance for two successive emissions is of
\begin{eqnarray}\label{covariance of two emissions}
\nonumber
\delta E'^{2(\mathrm{cov})}_\mathrm{ECNT}(M)&=&
\langle E'_1(M)E'_2(M)\rangle_\mathrm{ECNT}-
\langle E'_1(M)\rangle_\mathrm{ECNT}\langle E'_2(M)\rangle_\mathrm{ECNT}
\\ \nonumber
&=&\frac{8}{3}\pi\left(4\pi M^2-1+\frac{\pi M(8M-\mathrm{Er}[2\sqrt{\pi}M])}{1-e^{4\pi M^2}+2\pi M\mathrm{Er}[2\sqrt{\pi}M]}\right)
\\
&&-\frac{\big(4-4e^{4\pi M^2}+8\pi M\mathrm{Er}[2\sqrt{\pi}M]\big)^2}{\mathrm{Er}^2[2\sqrt{\pi}M]},
\end{eqnarray}
which has two limits,
\begin{eqnarray}\label{limits1}
&&\delta E'^{2(\mathrm{cov})}_\mathrm{ECNT}(M\rightarrow0)\sim-\frac{32\pi^2M^2}{3}+\frac{96\pi^3M^4}{5}+...,
\\ \label{limits2}
&&\delta E'^{2(\mathrm{cov})}_\mathrm{ECNT}(M\rightarrow\infty)
\sim-\frac{29}{16\pi M^4}.
\end{eqnarray}
Let's note that Eq. (\ref{covariance of two emissions}) is the same with that for the non-corrected non-thermal spectrum obtained in Ref. \cite{zhang5}.

\begin{figure}[ht]
\centering
\includegraphics[width=0.46\textwidth]{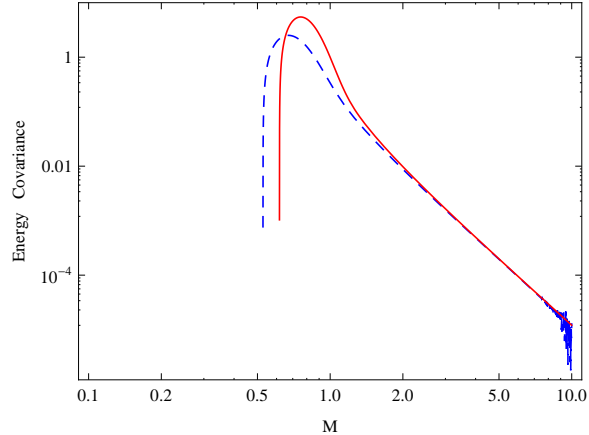}
\caption{(color online) The covariance of successive emissions is nontrivial for the non-thermal spectra. The red line represents the temperature-corrected case, and the blue dashed one represents the energy-corrected case.}\label{f9}
\end{figure}

We can see from Fig. \ref{f9} that the covariances approaches their maximums also near the Planck mass scale, no matter for the temperature-corrected case or the energy-corrected case. However, they have different maximums, and the maximum covariance for the temperature-corrected non-thermal spectrum is bigger than that of the energy-corrected non-thermal one. Furthermore, it is interesting that both the covariances vanish at small or large masses. As discussed in Fig. \ref{f7}, for both the quantum corrected spectra their average number of emissions become limited (for instance, two emissions) when the mass is small. Thus, the covariances vanish at the small mass limit. Which is consistent with the approximate analytical result for the energy-corrected non-thermal spectrum given in Eq. (\ref{limits1}) when $M\ll 1$. For large masses, the covariances decrease quickly, and finally approach zero for extremely large mass. This results from the sharp decrease of the average emission energies illustrated in Fig. \ref{f6}. Furthermore, the correlation between the two emissions is proportional to their product. As a result of that, it is reasonable that the covariances decrease at the large mass limit.


\section{Discussions and Conclusions} \label{section 6}

We investigate the long-standing ``information loss paradox" and
the features of the non-corrected thermal (non-thermal) spectrum and the temperature-corrected (energy-corrected) thermal (non-thermal) spectrum.
Our analysis show that the largest covariances of successive emissions appear near the Planck mass scale, i.e., for the temperature-corrected non-thermal spectrum $\delta E^{2(\mathrm{cov})}_\mathrm{TCNT}\sim5.4$ for $M\sim0.77M_P$ , and for the energy-corrected non-thermal spectrum $\delta E^{2(\mathrm{cov})}_\mathrm{ECNT}\sim2.5$ for $M\sim0.67M_P$~\footnote{For $D$ dimensional Schwarzschild black hole, the fundamental Planck scale is reduced depending on the compact space of volume $V_{D-4}$, e.g., the reduced Planck scale $M_P\sim1\mathrm{TeV}$ with $D=10$ and $V_6\sim fm^6$ \cite{Giddings}}. Here we give an open problem that the possibility of different radiation spectra discussed above may also exist for other black hole systems, such as the micro black holes model discussed extensively in connection with the experiments of the CERN large Hadron Collider (LHC) \cite{Emparan,Dimopoulos,Giddings,Meade,CMS1}. According to research \cite{CMS1}, it is estimated that the minimum black hole mass should be in the range of $3.5-4.5~\mathrm{TeV}$ for $pp$ collisions at a center-of-mass energy of $7~\mathrm{TeV}$ at LHC. Based on it, a more recent study \cite{CMS2} showed that the limits on the minimum semiclassical black hole and string-ball masses in the range $3.8$ to $5.3\mathrm{TeV}$ for a wide range of model parameters. So, if the radiation of a micro black hole were observed, then it may be possible to use the energy covariances as a indicator to determine whether the emission spectrum is non-thermal or not, and whether the emission spectrum is temperature-corrected or not. Also note that the energy scale about the production and observation of micro black holes is being debated \cite{Mureika}. Thus, when and whether the micro black holes could be observed on Earth, especially in a LHC experiment, needs a lot of work. On the other hand, other kinds of manmade black holes, such as those implemented or discussed with optical, acoustic, and cold-atomic systems \cite{Un,Belgiorno,Rousseaux,Weinfurtner,Lahav}, are being discussed,
and several experiments \cite{Belgiorno,Rousseaux,Weinfurtner,Lahav} had shown evidence of Hawking radiation from the event horizon. Therefore, the same problems we discussed above are also worthy of being studied in these real radiation systems.

We have discussed several radiation spectra of Schwarzschild black hole, the non-corrected thermal (non-thermal) spectrum, the temperature-corrected thermal (non-thermal) spectrum, and the energy-corrected thermal (non-thermal) spectrum. We emphasize that the non-thermal property of radiation spectra comes from taking account of the energy conservation during the evaporation process, and the quantum corrections results from the Hamilton-Jacobi method beyond semiclassical approximation. Furthermore, we use the temperature-corrected non-thermal spectrum and the energy-corrected one to solve the long-standing ``information loss paradox" of black hole. We find that the entropy of the emitted particles, which exhaust the initial black hole, is identical to the original entropy of the black hole, thus entropy is conserved between the initial (black hole plus no radiation) and final (no black hole plus radiation field quanta) states. Which reveals that no information is lost, and the black hole evaporation process in unitary. These discussions, therefore, may provide a possible way to understand the ``information loss paradox".

To distinguish different radiation spectra, their corresponding average emission energies, average numbers of emissions and average emission energy fluctuations are compared. It is found that there are no obvious differences between them except for near the Planck mass scale. Especially, the energy-corrected spectra have the same corresponding average emission energies, average numbers of emissions and average emission energy fluctuations  with that for the non-corrected spectra. Of great interest, we find that for all the thermal spectra the energy covariances of Hawking radiations completely vanish, while they are nontrivial for all the non-thermal spectra. Especially, the temperature-corrected non-thermal spectrum and the energy-corrected one have distinctly different maximums of energy covariances. As a result of that, these differences provide a way towards experimentally studying the long-standing puzzles of Hawking radiation that whether the radiation spectrum of black hole is thermal or non-thermal with or without high order quantum corrections.

\begin{acknowledgments}
This work was supported by the National Natural Science Foundation of China under Grant Nos. 11175065, 10935013; the National Basic Research of China under Grant No. 2010CB833004; the Hunan Provincial Natural Science Foundation of China under Grant No. 11JJ7001; Hunan Provincial Innovation Foundation For Postgraduate under Grant No CX2012B202; the Construct Program of the National  Key Discipline.

\end{acknowledgments}

\end{document}